\documentclass[twocolumn,aps,floatfix,showpacs,showkeys,prb]{revtex4}
\usepackage{epsfig}
\usepackage{graphicx}
\usepackage{amsmath}
\usepackage{amssymb}
\usepackage{amsfonts}
\usepackage{tabularx}
\usepackage{color}
\usepackage{dcolumn}
\usepackage{bm}

\begin{document}

\title{First principles theory of  chiral dichroism in electron microscopy applied to 3d ferromagnets}

\author{J\'{a}n Rusz} \email{rusz@mag.mff.cuni.cz}
\affiliation{Institute of Physics, Academy of Sciences of the Czech Republic, 
Cukrovarnick\'{a} 10, 162~53 Prague 6, Czech Republic}
\altaffiliation[currently at ]{Department of Physics, Uppsala University, Box 530, S-751~21 Uppsala, Sweden}

\author{Stefano Rubino}
\author{Peter Schattschneider}
\affiliation{Institute for Solid State Physics, Vienna University of Technology, Wiedner Hauptstrasse 8-10/138, A-1040 Vienna, Austria}


\date{\today}

\begin{abstract}
Recently it was demonstrated (Schattschneider {\it et al.}, Nature \textbf{441} (2006), 486), that an analogue of the X-ray magnetic circular dichroism (XMCD) experiment can be performed with the transmission electron microscope (TEM). The new phenomenon has been named energy-loss magnetic chiral dichroism (EMCD). 
In this work we present a detailed {\it ab initio} study of the chiral dichroism in the Fe, Co and Ni transition elements. We discuss the methods used for the simulations together with the validity and accuracy of the treatment, which can, in principle, apply to any given crystalline specimen. The dependence of the dichroic signal on the sample thickness, accuracy of the detector position and the size of convergence and collection angles is calculated.
\end{abstract}


\maketitle

\section{Introduction}

The analogy between X-ray absorption spectroscopy (XAS) and electron energy loss spectroscopy (EELS) has been recognized long ago~\cite{Hitchcock,Yuan}. The role of the polarization vector $\bm{\varepsilon}$ in XAS is similar to the role of the wave vector transfer $\mathbf{q}$ in EELS. 
This has made feasible the detection of linear dichroism in the TEM. However the counterpart of X-ray magnetic {\em circular} dichroism (XMCD)~\cite{Thole,Lovesey,Stohr95} experiments with electron probes was thought to be  technically impossible due to the low intensity of existing spin polarized electron sources.
XMCD is an important technique providing atom-specific information about the magnetic properties of materials. Particularly the near edge spectra, where a well localized strongly bound electron with $l \ne 0$ is excited to an unoccupied band state, allow to measure spin and orbital moments. 
Soon after the proposal of an experimental setup for detection of circular dichroism using a standard non-polarized electron beam in the TEM~\cite{HebertUM03}  it was demonstrated that such experiments (called energy-loss magnetic chiral dichroism, EMCD) are indeed possible\cite{nature}. 
This novel technique is of considerable interest for nanomagnetism and spintronics according to the high spatial resolution of the TEM.  However, its  optimization involves many open questions.

In this work we provide theoretical {\it ab initio} predictions of the dependence of the dichroic signal in the EMCD experiment on several experimental conditions, such as sample thickness, detector placement and the finite size of convergence and collection angles. This information should help to optimize the experimental geometry in order to maximize the signal to noise ratio.

The structure of this work is as follows: In section II we first describe the computational approach based on the dynamical diffraction theory and electronic structure calculations. We also discuss the validity of several approximations for the mixed dynamic form factor. In section III we study the dependence of the dichroic signal of bcc-Fe, hcp-Co and fcc-Ni on various experimental conditions. This section is followed by a concluding section summarizing the most important findings.

\section{Method of calculation}

We will follow the derivations of the double differential scattering cross-section (DDSCS) presented in Refs.~\onlinecite{NelhiebelPRL00,Micha}. Within the first-order Born approximation~\cite{SchattschneiderPRB05} 
DDSCS is written as
\begin{equation} \label{Born}
  \frac{\partial^2 \sigma}{\partial\Omega \partial E} = \frac{4\gamma^2}{a_0^2} \frac{k_f}{k_0} \frac{S(\mathbf{q},E)}{q^4}
\end{equation}

with \begin{equation} \label{eqdff}
 S(\mathbf{q},E) = \sum_{i,f} |\langle i | e^{i\mathbf{q}\cdot\hat{\mathbf{R}}} | f \rangle|^2 \delta{(E_f - E_i - E)}
\end{equation} 
where $\mathbf{q} = \mathbf{k}_f - \mathbf{k}_0$ is the difference (wave vector transfer) between final 
wave vector $\mathbf{k}_f$ and initial wave vector $\mathbf{k}_0$ of the fast electron; 
$\gamma=1/\sqrt{1-v^2/c^2}$ is a relativistic
 factor and $a_0$ is Bohr radius. The $S(\mathbf{q},E)$ is the so called dynamic form factor (DFF)~\cite{KohlRose85}.

This equation is valid only if the initial and final wave functions of the fast electron are plane waves. In the crystal
the full translation symmetry is broken and as a result, the electron wave function becomes a superposition of Bloch waves, 
which reflects the discrete translation symmetry. Each Bloch wave can be decomposed into a linear combination of plane 
waves - it is a coherent superposition of (an in principle infinite number of) plane waves. The wave function of the fast electron can be thus written as
\begin{equation} \label{fewf}
  \psi(\mathbf{r}) = \sum_\mathbf{g} \sum_j \epsilon^{(j)} C_\mathbf{g}^{(j)} e^{i(\mathbf{k}^{(j)}+\mathbf{g}) \cdot \mathbf{r}}
\end{equation}
for incident wave and
\begin{equation} \label{fewfw}
  \psi'(\mathbf{r}) = \sum_\mathbf{h} \sum_l \epsilon^{(l)} D_\mathbf{h}^{(l)} e^{i(\mathbf{k}^{(l)}+\mathbf{h}) \cdot \mathbf{r}}
\end{equation}
for outgoing wave, where $C_\mathbf{g}^{(j)}$, $D_\mathbf{h}^{(l)}$ are so called Bloch coefficients, $\epsilon^{(j)}$ ($\epsilon^{(l)}$) determine the excitation of the Bloch wave with index $j$ ($l$) and wave vector $\mathbf{k}^{(j)}$ ($\mathbf{k}^{(l)}$) and $\mathbf{g}$ ($\mathbf{h}$) is a vector of the reciprocal lattice.

When we derive the Born approximation of DDSCS starting with such fast electron wave functions, we will obtain a sum of two kinds of terms: direct terms (DFFs) as in the plane wave Born approximation Eq.~(\ref{Born}), and interference terms. These interference terms are a generalization of the DFF - the mixed dynamic form factors\cite{KohlRose85} (MDFFs). Each of them is defined by two wave vector transfers, thus we label them $S(\mathbf{q},\mathbf{q'},E)$.

The MDFF can be evaluated within a single particle approximation as
\begin{equation} \label{mdff}
  S(\mathbf{q},\mathbf{q'},E) = \sum_{i,f} \langle i | e^{i\mathbf{q} \cdot \hat{\mathbf{R}}} | f \rangle
         \langle f | e^{-i\mathbf{q'} \cdot \hat{\mathbf{R}}} | i \rangle \delta{(E_f - E_i - E)}
\end{equation}
where $|i\rangle,|f\rangle$ are the initial and final single-electron wave functions of the target electron in the crystal.
Thus the definition of MDFF encompasses the notion of DFF, Eq.~(\ref{eqdff}), for $\mathbf{q}=\mathbf{q'}$.
For more details about calculation of MDFF see subsection \ref{sec:mdff}.

The wave vector transfers are 
$\mathbf{q}_{\mathbf{gh}}^{jl} = \mathbf{k}^{(l)}-\mathbf{k}^{(j)} + \mathbf{h}-\mathbf{g}$
and the total DDSCS will be a sum over all diads of $\mathbf{q}$ and $\mathbf{q'}$ vectors of terms
\begin{equation} \label{dscsterm}
  \frac{4\gamma^2}{a_0^2} \frac{\chi_f}{\chi_0} 
  \sum_\mathbf{a} X_{\mathbf{ghg}'\mathbf{h}'}^{jlj'l'}(\mathbf{a}) \frac{S_\mathbf{a}(\mathbf{q},\mathbf{q'},E)}{q^2 q'^2}
\end{equation}
where $X_{\mathbf{ghg}'\mathbf{h}'}^{jlj'l'}(\mathbf{a})$ is a the product of the coefficients of the individual plane wave
components of the fast electron wave functions and $\mathbf{a}$ labels the position of the atoms where the inelastic event can occur.
The $X_{\mathbf{ghg}'\mathbf{h}'}^{jlj'l'}(\mathbf{a})$ coefficients are given by dynamical diffraction theory. This
will be covered in the next subsection \ref{sub:ddt}. The $\chi_f$ and $\chi_0$ are magnitudes of wave vectors
outside the crystal (in the vacuum).

The calculation is thus split into two separate tasks. i) Calculation of Bloch wave coefficients using the
dynamical diffraction theory and identification of important terms. This task is mainly `geometry dependent', although
it can also contain some input from electronic structure codes, namely the Coulomb part of crystal potential.
ii) Calculation of MDFFs requested by the dynamical diffraction theory. This part strongly depends on the electronic structure of the studied system. The final step is the summation of all terms.

\subsection{Dynamical diffraction theory \label{sub:ddt}}

The formalism, which will be described here is a generalization of the formalism presented in Ref.~\onlinecite{Micha,nelhelnes} extending it beyond systematic row approximation by including also higher-order Laue zones (HOLZ). The extension to HOLZ is performed along lines presented in Refs.~\onlinecite{Metherell,PengWhelan}. We will assume the high-energy Laue case, {\it i.e.}\ we can safely neglect back-reflection and back-diffraction.

The Bloch wave vectors of the electron after entering the crystal fulfill the continuity condition
\begin{equation} \label{contin}
  \mathbf{k}^{(j)} = \mathbf{\chi} + \gamma^{(j)} \mathbf{n}
\end{equation}
where $\mathbf{n}$ is the unit vector normal to the crystal surface and $\chi$ is the wave vector of the incoming electron. Only the wave vector component normal to the surface can change.

Expanding the wave function of the fast electron into a linear combination of plane waves and substituting it into the Schr\"{o}dinger equation we obtain the secular equation~\cite{Metherell}
\begin{equation}
  \sum_\mathbf{g} \left[ \left( K^2 - (\mathbf{k}^{(j)}+\mathbf{g})^2 \right) + \sum_{\mathbf{h} \ne 0} U_\mathbf{h} C_{\mathbf{g}-\mathbf{h}}^{(j)} \right] e^{i(\mathbf{k}^{(j)}+\mathbf{g}) \cdot \mathbf{r}}=0
\end{equation}
where $K^2 = U_0 + 2meE/\hbar^2$, $m$ and $e$ are, respectively, the electron mass and charge,
$U_\mathbf{g}=2meV_\mathbf{g}/\hbar^2$ where $V_\mathbf{g}$ are
the Fourier components of the crystal potential, which can
be either calculated \textit{ab initio} \cite{wien2k,vgnote} or obtained from the tabulated forms of the potential~\cite{WeickKohl,DoyleTurner}.
 It can be shown \cite{Lewis78,Metherell} that in the high energy limit the secular equation, 
which is a quadratic eigenvalue problem in $\gamma^{(j)}$, can be reduced to a linear eigenvalue problem 
$\mathbb{AC}^{(j)} = \gamma^{(j)}\mathbb{C}^{(j)}$ where $\mathbb{A}$ is a non-hermitean matrix\cite{Lewis78,PengWhelan}

\begin{equation}
  A_{\mathbf{gh}} = \frac{K^2-(\mathbf{\chi}+\mathbf{g})^2}{2(\mathbf{\chi}+\mathbf{g}) \cdot \mathbf{n}} \delta_{\mathbf{gh}} +
                    (1-\delta_{\mathbf{gh}}) \frac{U_{\mathbf{g}-\mathbf{h}}}{2(\mathbf{\chi}+\mathbf{g}) \cdot \mathbf{n}}
\end{equation}

 This eigenvalue problem can be transformed into a hermitean one using a diagonal matrix
$\mathbb{D}$ with elements
\begin{equation}
  \mathbb{D}_{\mathbf{gh}} = \delta_{\mathbf{gh}} \left[ 1 + \frac{\mathbf{g}\cdot \mathbf{n}}{\mathbf{\chi} \cdot \mathbf{n}} \right]
\end{equation}
Then the eigenvalue problem is equivalent to $(\mathbb{D}^{1/2}\mathbb{AD}^{-1/2})(\mathbb{D}^{1/2}\mathbb{C}^{(j)}) = 
\gamma^{(j)} (\mathbb{D}^{1/2}\mathbb{C}^{(j)})$ or $\tilde{\mathbb{A}}\tilde{\mathbb{C}}^{(j)} = 
\gamma^{(j)}\tilde{\mathbb{C}}^{(j)}$, where the matrix $\tilde{\mathbb{A}}$ is hermitean

\begin{eqnarray}
  \tilde{A}_{\mathbf{gh}} & = & 
       \frac{K^2-(\mathbf{\chi}+\mathbf{g})^2}{2(\mathbf{\chi}+\mathbf{g}) \cdot \mathbf{n}} \delta_{\mathbf{gh}} + \\
                          & + & (1-\delta_{\mathbf{gh}}) \frac{U_{\mathbf{g}-\mathbf{h}}}
       {2 \sqrt{ [(\mathbf{\chi}+\mathbf{g}) \cdot \mathbf{n}] [(\mathbf{\chi}+\mathbf{h}) \cdot \mathbf{n}]}} \nonumber 
\end{eqnarray}
and the original Bloch wave coefficients can be retrieved using the relation

\begin{equation} \label{Bloch}
  C^{(j)}_\mathbf{g} = \tilde{C}^{(j)}_\mathbf{g} 
        \left/ \sqrt{ 1 + \frac{\mathbf{g} \cdot \mathbf{n}}{\mathbf{\chi} \cdot \mathbf{n}} } \right.
\end{equation}

By solving this eigenvalue problem we obtain the fast electron wave function as a linear combination of eigenfunctions as given
in Eq.~(\ref{fewf}). To obtain values for $\epsilon^{(j)}$ we need to impose boundary conditions, namely that the electron is
described by a single plane wave at the crystal surface. The crystal surface is a plane defined by the scalar product 
$\mathbf{n}\cdot\mathbf{r}=t_0$. Then the boundary condition (in the high energy limit) leads to the following condition\cite{Metherell}
\begin{equation}
  \epsilon^{(j)} = C_{\mathbf{0}}^{(j)\star} e^{-i\gamma^{(j)} t_0}
\end{equation}
It is easy to verify that
\begin{eqnarray}
  \lefteqn{\psi(\mathbf{r})|_{\mathbf{n}\cdot\mathbf{r}=t_0} = } \nonumber \\
  & = & \sum_{j\mathbf{g}} 
     C_{\mathbf{0}}^{(j)\star} C_{\mathbf{g}}^{(j)} e^{i(\mathbf{k}^{(j)}+\mathbf{g}) \cdot \mathbf{r}} e^{-i\gamma^{(j)} t_0}
     \nonumber \\
  & = & \sum_{\mathbf{g}} e^{i(\mathbf{\chi}+\mathbf{g}) \cdot \mathbf{r}} 
  \sum_j e^{i\gamma^{(j)}\mathbf{n}\cdot\mathbf{r}} e^{-i\gamma^{(j)}t_0} C_{\mathbf{0}}^{(j)\star} C_{\mathbf{g}}^{(j)}
     \nonumber \\
  & = & \sum_{\mathbf{g}} e^{i(\mathbf{\chi}+\mathbf{g})\cdot\mathbf{r}} 
  \sum_j C_{\mathbf{0}}^{(j)\star} C_{\mathbf{g}}^{(j)}
     \nonumber \\
  & = & \sum_\mathbf{g} e^{i(\mathbf{\chi}+\mathbf{g})\cdot\mathbf{r}} \delta_{\mathbf{0g}} 
  \left/ \sqrt{ 1 + \frac{\mathbf{g}\cdot\mathbf{n}}{\mathbf{\chi}\cdot\mathbf{n}} } \right.
     \nonumber \\
  & = &
        e^{i\mathbf{\chi}\cdot\mathbf{r}} |_{\mathbf{n}\cdot\mathbf{r}=t_0}
\end{eqnarray}
as required by the boundary condition. We have used the continuity condition, Eq.~(\ref{contin}), and the 
completeness relation for the Bloch coefficients
\begin{eqnarray}
  \delta_{\mathbf{gh}} & = & \sum_j \tilde{C}_\mathbf{g}^{(j)*} \tilde{C}_\mathbf{h}^{(j)} = \\
                       & = &
  \sqrt{\left[ 1 + \frac{\mathbf{g}\cdot\mathbf{n}}{\mathbf{\chi}\cdot\mathbf{n}} \right]
        \left[ 1 + \frac{\mathbf{h}\cdot\mathbf{n}}{\mathbf{\chi}\cdot\mathbf{n}} \right]} \nonumber 
        \sum_j C_\mathbf{g}^{(j)*} C_\mathbf{h}^{(j)}
\end{eqnarray}

Therefore the wave function of the fast moving electron in the crystal, which becomes a single plane wave at $\mathbf{n}\cdot\mathbf{r}=t_0$ is given
by the following expression
\begin{equation}
  \psi(\mathbf{r}) = \sum_{j\mathbf{g}} 
       C_{\mathbf{0}}^{(j)\star} C_{\mathbf{g}}^{(j)} e^{i\gamma^{(j)}(\mathbf{n}\cdot\mathbf{r}-t_0)} e^{i(\mathbf{\chi}+\mathbf{g})\cdot\mathbf{r}}
\end{equation}

The following discussion will be restricted to a particular case - a crystal with parallel surfaces.
For such a crystal with normals in the direction of the $z$ axis we set
$t_0=0$ for the fast electron entering the crystal and $t_0=t$ when leaving the crystal ($t$ is the crystal thickness).

The inelastic event leads to a change of the energy and momentum of the scattered electron. The detector position determines the observed projection of the electron wave function (Bloch field) onto a plane wave after the inelastic event. Therefore the calculation of the ELNES requires the solution of two independent eigenvalue problems describing an electron wave function before and after the inelastic event\cite{Micha,nelhelnes}. Invoking reciprocity for electron propagation the outgoing wave can also be considered as a time reversed solution of the Schr\"odinger equation, also known as the {\it reciprocal wave}\cite{Kainuma} with the source replacing the detector position.  

Now we can identify the prefactors $X_{\mathbf{ghg}'\mathbf{h}'}^{jlj'l'}(\mathbf{a})$ from Eq.~(\ref{dscsterm}). 
For the sake of clarity we will keep $C_\mathbf{g}^{(j)}$ for the Bloch coefficients of the incoming electron and we 
use $D_\mathbf{h}^{(l)}$ for the Bloch coefficients of the outgoing electron entering the detector
(obtained from the two independent eigenvalue problems). Similarly, superscript indices $(j)$ and $(l)$ indicate
eigenvalues and Bloch-vectors for incoming and outgoing electron, respectively. We thus obtain
\begin{eqnarray}
  X_{\mathbf{ghg}'\mathbf{h}'}^{jlj'l'}(\mathbf{a}) & = &
   C_{\mathbf{0}}^{(j)\star} C_{\mathbf{g}}^{(j)} 
   D_{\mathbf{0}}^{(l)} D_{\mathbf{h}}^{(l)\star} 
                       \nonumber \\  & \times &
   C_{\mathbf{0}}^{(j')} C_{\mathbf{g'}}^{(j')\star}
   D_{\mathbf{0}}^{(l')\star} D_{\mathbf{h'}}^{(l')}
                       \nonumber \\  & \times &
   e^{i(\gamma^{(l)}-\gamma^{(l')})t} e^{i(\mathbf{q}-\mathbf{q'})\cdot\mathbf{a}}
\end{eqnarray}
where
\begin{eqnarray}
  \mathbf{q}  & = & \mathbf{k}^{(l)}-\mathbf{k}^{(j)} + \mathbf{h}-\mathbf{g} \nonumber \\
  \mathbf{q'} & = & \mathbf{k}^{(l')}-\mathbf{k}^{(j')} + \mathbf{h'}-\mathbf{g'}
\end{eqnarray}

In crystals the position of each atom can be decomposed into a sum of a lattice vector and a base vector, $\mathbf{a} = \mathbf{R} + \mathbf{u}$. Clearly, MDFF does not depend on $\mathbf{R}$, but only on $\mathbf{u}$. It is then possible to perform
analytically the sum over all lattice vectors $\mathbf{R}$ under the approximation that the MDFF does not depend strongly on 
the $j,l$ indices. This is indeed a very good approximation, as verified by numerical simulations (see below).

First we will treat the summation over all lattice vectors. The sum in Eq.~\ref{dscsterm} can be separated into two terms
\begin{eqnarray}
  \frac{1}{N} \sum_{\mathbf{a}} e^{i(\mathbf{q}-\mathbf{q'})\cdot\mathbf{a}} & = &
  \frac{1}{N_\mathbf{u}} \sum_{\mathbf{u}} e^{i(\mathbf{q}-\mathbf{q'})\cdot\mathbf{u}}
  \frac{1}{N_\mathbf{R}} \sum_{\mathbf{R}} e^{i(\mathbf{q}-\mathbf{q'})\cdot\mathbf{R}}
\end{eqnarray}
Since 
\begin{eqnarray} \label{qmqp}
  \mathbf{q}-\mathbf{q'} & = & [(\gamma^{(j)}-\gamma^{(j')})-(\gamma^{(l)}-\gamma^{(l')})] \mathbf{n} \nonumber \\
                         & + & \mathbf{h}-\mathbf{h'}+\mathbf{g'}-\mathbf{g}
\end{eqnarray}
and the algebric sum of $\mathbf{g},\mathbf{h}$ is simply
 a reciprocal lattice vectors $\mathbf{G}$, which fulfills $e^{i\mathbf{G}\cdot\mathbf{R}}=1$, it is possible to simplify
the second term
\begin{equation}
  \sum_{\mathbf{R}} e^{i(\mathbf{q}-\mathbf{q'})\cdot\mathbf{R}} = 
  \sum_{\mathbf{R}} e^{i[(\gamma^{(j)}-\gamma^{(j')})-(\gamma^{(l)}-\gamma^{(l')})] \mathbf{n}\cdot\mathbf{R}}
\end{equation}

For general orientations of the vector $\mathbf{n}$ this sum is difficult to evaluate. In particular coordinate
system with $\mathbf{n} \parallel z$ and crystal axes $a,b \perp z$ this sum leads \cite{Micha,Schattelspec2005} to
\begin{equation}
  \sum_{\mathbf{R}} e^{i(\mathbf{q}-\mathbf{q'})\cdot\mathbf{R}} = 
  N_\mathbf{R} e^{i\Delta t/2} \frac{\sin \Delta \frac{t}{2}}{\Delta \frac{t}{2}}
\end{equation}
so that the total sum over all atomic positions is
\begin{eqnarray}
  \frac{1}{N} \sum_{\mathbf{a}} e^{i(\mathbf{q}-\mathbf{q'})\cdot\mathbf{a}} & = &
  e^{i\Delta \frac{t}{2}} \frac{\sin \Delta \frac{t}{2}}{\Delta \frac{t}{2}}
  \frac{1}{N_\mathbf{u}} \sum_{\mathbf{u}} e^{i(\mathbf{q}-\mathbf{q'})\cdot\mathbf{u}}
\end{eqnarray}
where $\Delta = (\gamma^{(j)}-\gamma^{(j')})-(\gamma^{(l)}-\gamma^{(l')})$. The final expression of the DDSCS we write as
\begin{eqnarray} \label{dscsfin}
  \frac{\partial^2 \sigma}{\partial\Omega \partial E} & = & \sum_{\mathbf{ghg}'\mathbf{h}'} 
    \frac{1}{N_\mathbf{u}} \sum_\mathbf{u}  
                                           \frac{S_\mathbf{u}(\mathbf{q},\mathbf{q'},E)}{q^2 q'^2} e^{i(\mathbf{q}-\mathbf{q'})\cdot\mathbf{u}}
                      \nonumber \\  & \times & 
    \sum_{jlj'l'} Y_{\mathbf{ghg}'\mathbf{h}'}^{jlj'l'} T_{jlj'l'}(t)
\end{eqnarray}
where
\begin{eqnarray}
  Y_{\mathbf{ghg}'\mathbf{h}'}^{jlj'l'} & = &
   C_{\mathbf{0}}^{(j)\star} C_{\mathbf{g}}^{(j)}
   D_{\mathbf{0}}^{(l)} D_{\mathbf{h}}^{(l)\star}
                       \\  & \times &
   C_{\mathbf{0}}^{(j')} C_{\mathbf{g'}}^{(j')\star}
   D_{\mathbf{0}}^{(l')\star} D_{\mathbf{h'}}^{(l')}
                       \nonumber
\end{eqnarray}
depends only on the eigenvectors of the incoming and outgoing beam and
\begin{equation}
  T_{jlj'l'}(t) = e^{i[(\gamma^{(j)}-\gamma^{(j')})+(\gamma^{(l)}-\gamma^{(l')})]\frac{t}{2}} 
                      \frac{\sin \Delta \frac{t}{2}}{\Delta \frac{t}{2}}
\end{equation}
is a thickness and eigenvalue dependent function.

Perturbative treatment of the absorption can be easily introduced. If we denote by $U'_\mathbf{g}$ the absorptive part of the potential, within the first order perturbation theory the Bloch coefficients will not change, just the eigenvalues will be shifted by $i\eta^{(j)}$ or $i\eta^{(l)}$ for the incoming or outgoing wave, respectively. Particular $\eta^{(j)}$ can be calculated using the following expression\cite{Metherell}
\begin{equation}
  \eta^{(j)} = \frac{\sum_{\mathbf{g,h}}U'_\mathbf{g-h}C^{(j)}_\mathbf{h}C^{(j)\star}_\mathbf{g}}{2\sum_\mathbf{g} C^{(j)}_\mathbf{g} C^{(j)\star}_\mathbf{g}(\mathbf{\chi}+\mathbf{g})\cdot\mathbf{n}}
\end{equation}
and similarly for the outgoing beam.

This way the eigenvalues change from $\gamma^{(j)}$ to $\gamma^{(j)}+i\eta^{(j)}$ and the $\Delta$ acquires an imaginary part. Such approximative treatment of absorption thus affects only the thickness-dependent function $T_{jlj'l'}(t)$.

Here we add a few practical considerations, which we applied in our computer code. The sum in Eq.~(\ref{dscsfin}) is performed over 8 indices for every energy and thickness value. Such summation can easily grow to a huge number of terms and go beyond the computational capability of modern desktop computers. For example, if we assume the splitting of the incoming (and outgoing) beam into only 10 plane wave components, taking into account the 10 most strongly excited Bloch waves, we would have $10^8$ terms per each energy and thickness. A calculation with an energy mesh of 100 points at 100 different thicknesses would include one trillion terms and require a considerable amount of computing time. However most of these terms give a negligible contribution to the final sum. Therefore several carefully chosen cut-off conditions are required to keep the computing time reasonable without any significant degradation of the accuracy.

The first cut-off condition used is based on the Ewald's sphere construction. Only plane wave components with $\mathbf{k}+\mathbf{g}$ close 
to the Ewald's sphere will be excited. The strength of the excitation decreases also with decreasing crystal potential component
$U_\mathbf{g}$. A dimensionless parameter $w_\mathbf{g} = s_\mathbf{g} \xi_\mathbf{g}$ - product of the excitation error
and the extinction distance\cite{Metherell} - reflects both these criteria. Therefore we can 
filter the list of beams by selecting only beams 
with $w_\mathbf{g} < w_{max}$. Experience shows that in the final summation a fairly low number of beams is 
necessary to have a well converged results (in systematic row conditions this number is typically around 10).
The convergence of the corresponding Bloch coefficients
requires solving an eigenvalue problem with a much larger set of beams (several hundreds). Therefore we defined two
cut-off parameters for $w_\mathbf{g}$ - the first for the solution of the eigenvalue problem (typically $w_{max,1}$ is
between $1000$ and $5000$) and the second for the summation ($w_{max,2}$ typically between $50$ and $100$).

The second type of cut-off conditions is applied to selection of Bloch waves, which enter the summation. Once the set of
beams for summation is determined, this amounts to sorting the Bloch waves according to a product of their
excitation $\epsilon^{(j)}$ and their norm on the subspace defined by selected subset of beams, 
$C_\mathbf{0}^{(j)} || \mathbf{C}^{(j)} ||_{\text{subsp}}$. In the systematic row conditions this value is large only
for a small number of Bloch waves. Typically in the experimental geometries used for detection of EMCD one can perform a 
summation over less than 10 Bloch waves to have a well converged result (often 5 or 6 Bloch 
waves are enough).

\subsection{Mixed dynamic form factor \label{sec:mdff}}

\begin{figure}[t]
  \includegraphics[width=7cm]{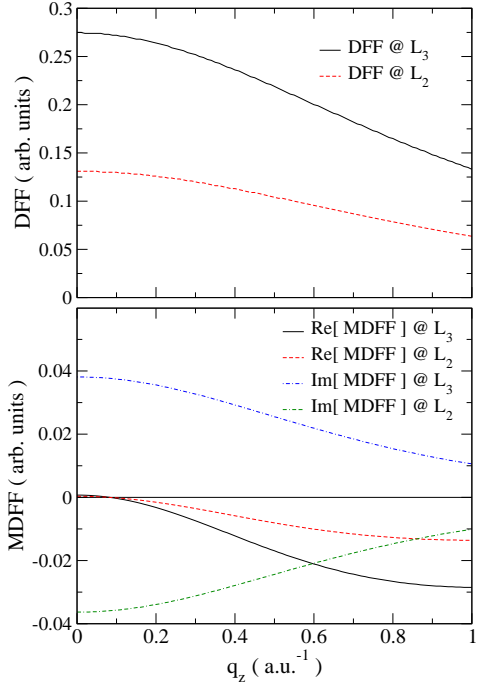}
  \caption{Dependence of $S(\mathbf{q},E)$ (top) and $S(\mathbf{q},\mathbf{q'},E)$ with $\mathbf{q'}=\mathbf{G}+\mathbf{q}$ (bottom) on $\mathbf{q}_z$, calculated for the $L_{2,3}$ edge of hcp-Co, with $\mathbf{G}=(100)$, $q_x=-q_x'=-|G|/2$, $q_y=q_y'=|G|/2$. 
  The ratio between values calculated at $L_3$ or $L_2$ is constant and equal to $2.1$ for the real part and to $-1$ for the imaginary part.\label{fig:mdffqz}}
\end{figure}

\begin{figure}[t]
  \includegraphics[width=7.5cm]{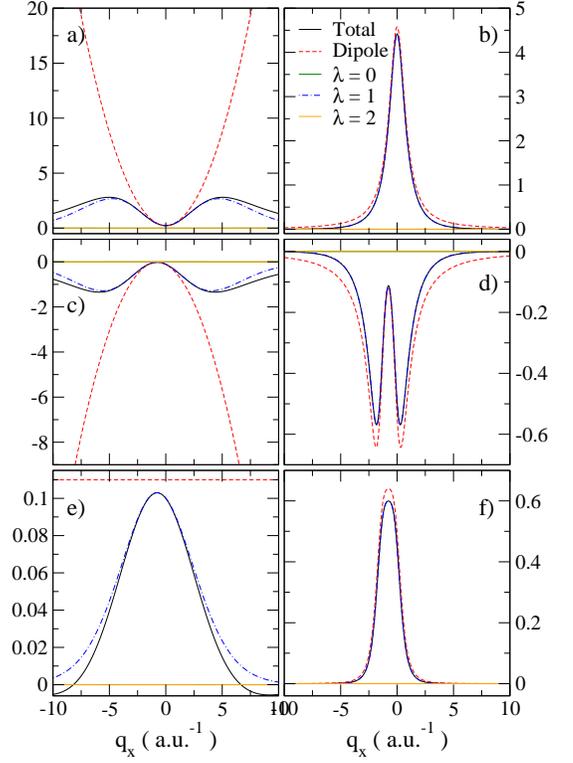}
  \caption{(online color) Decomposition of MDFF and dipole approximation calculated for hcp-Co with
           $\mathbf{q'}-\mathbf{q}=\mathbf{G}=(100)$ and $q_y = q'_y = |G|/2$ as a function of $q_x$ @ $L_3$. 
           Left column - graphs a), c) and e), show $S(\mathbf{q},\mathbf{q'},E)$ and 
           right column, graphs b), d) and f), show $S(\mathbf{q},\mathbf{q'},E)/q^2 q'^2$. 
           Top row - a) and b) - is the DFF,
           middle row - c) and d) - is the real part of MDFF and 
           bottom row - e) and f) - is the imaginary part of MDFF. 
           The $y$-axes are in arbitrary units, but consistent within the given column. The values for the $L_2$ edge differ only by a factor of $2.1$ for the real part and $-1$ for the imaginary part.
           Note that the contributions of $\lambda=0$ and $2$ are always negligible. See text for more details.
           \label{fig:dipapprox}}
\end{figure}

It can be seen from Eq.~(\ref{mdff}) that the calculation of the MDFF requires the evaluation of two matrix elements between initial and final states of the target electron. The derivation of the expression for the MDFF describing a transition from core state $nl\kappa$ ($n,l,\kappa$ are the main, orbital and relativistic quantum numbers, respectively) to a band state with energy $E$ is presented in detail in the supplementary material of Ref.~\onlinecite{nature} and in Ref.~\onlinecite{Micha}. Though, note that in Ref.~\onlinecite{Micha} the initial states are treated classically, which leads to somewhat different expression for MDFF giving incorrect $L_2-L_3$ branching ratio.

The final expression is\cite{nature} 
\begin{eqnarray*}\label{mdffbig}
    \lefteqn{S(\mathbf{q},\mathbf{q'},E) = }\\
     & = & \sum_{mm'} \sum_{LMS} \sum_{L'M'S'} \sum_{\lambda\mu} \sum_{\lambda'\mu'}  4\pi i^{\lambda-\lambda'} (2l+1) \sqrt{[\lambda,\lambda',L,L']} \\
 & \times & Y_\mu^\lambda(\mathbf{q}/q)^* Y_{\mu'}^{\lambda'}(\mathbf{q'}/q') \langle j_\lambda(q) \rangle_{ELSj} \langle j_{\lambda'}(q') \rangle_{EL'S'j} \\
 & \times &
\begin{pmatrix} l & \lambda  & L  \\  0  & 0    & 0  \end{pmatrix}
\begin{pmatrix} l & \lambda' & L' \\  0  & 0    & 0  \end{pmatrix}
\begin{pmatrix} l & \lambda  & L  \\ -m  & \mu  & M  \end{pmatrix}
\begin{pmatrix} l & \lambda' & L' \\ -m' & \mu' & M' \end{pmatrix} \\
 & \times & \sum_{j_z} (-1)^{m+m'} (2j+1)
\begin{pmatrix} l & \frac{1}{2} & j \\ m  & S  & -j_z \end{pmatrix}
\begin{pmatrix} l & \frac{1}{2} & j \\ m' & S' & -j_z \end{pmatrix} \\
 & \times & \sum_{\nu\mathbf{k}} D_{LMS}(\nu\mathbf{k}) D_{L'M'S'}(\nu\mathbf{k})^* \delta(E+E_{nl\kappa}-E_{\nu\mathbf{k}})
\end{eqnarray*}
Here we made use of Wigner 3j-symbols, $Y_\mu^\lambda$ are spherical harmonics, $\langle j_\lambda(q) \rangle_{ELSj}$ are radial integrals of all the radial-dependent terms (radial part of the wave function of the core and band states, radial terms of the Rayleigh expansion) and $D_{LMS}(\nu\mathbf{k})$ is the projection of the $(\nu\mathbf{k})$ Bloch state onto the $LMS$ subspace within the atomic sphere of the excited atom. For more details we refer to the supplementary material of Ref.~\onlinecite{nature}.

For evaluation of the radial integrals and Bloch state projections $D_{LMS}(\nu\mathbf{k})$ we employ the density functional theory\cite{dft} within the local spin density approximation\cite{lsda}.

In  Section~\ref{sub:ddt} we used an approximation of negligible dependence of MDFF on the $j,l$ indices (see Eq.\ref{qmqp}). Generally, as the wave vector $\mathbf{k}^{(j,l)}$ for each Bloch wave changes slightly by an amount given by the corresponding eigenvalue $\gamma^{(j,l)}$, the values of $q_z$ and $q'_z$ would change accordingly and therefore we should not be allowed to take MDFF out of the sum over the indices $j,l$ in the Eq.~(\ref{dscsfin}). However, the change in $q_z$ (and $q'_z$) induced by the eigenvalues $\gamma^{(j,l)}$ is small and can be neglected with respect to the $q_z=\chi_0 E/2 E_0$ given by the energy loss $E$\footnote{The question of momentum conservation in the $z$ direction in the inelastic interaction in a crystal of finite thickness is related to the probability of inter- and intrabranch transitions of the probe electron\cite{YoungRez}}. 
To demonstrate this we plot the dependence of MDFF on $q_z$, $q'_z$ for $q_x$ and $q_y$ corresponding to the main DFF and MDFF terms, see Fig.~\ref{fig:mdffqz}. If $q_z$ is given in a.u.$^{-1}$ (atomic units, 1 a.u.$=0.529178$\AA), typical values for L$_{2,3}$ edges of Fe, Co and Ni are around tenth of a.u.$^{-1}$, whereas typical values of $\gamma^{(j,l)}$ for strongly excited Bloch waves are one or two orders of magnitude smaller. Thus the approximation of weak $j,l$ dependence of MDFF is well justified.

\begin{figure*}[t]
  \includegraphics[width=12cm,angle=270]{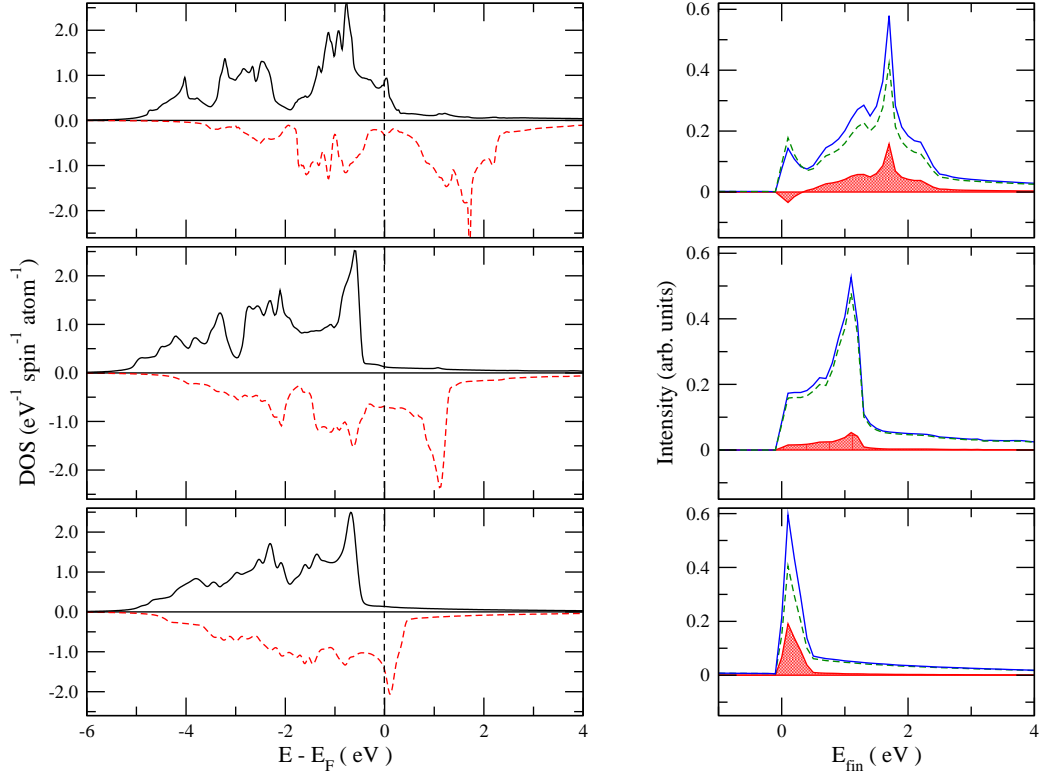}
  \caption{(online color) Spin-resolved $d$-densities of states (left) and resulting signal on $L_3$ edge (right) on bcc-Fe, hcp-Co and fcc-Ni (from top to bottom) at optimal thickness (see text). Spin-up DOS is drawn using a solid black line (positive) and spin-down DOS using a dashed red line (negative). DDSCS for the (+) detector position is drawn using a solid blue line, DDSCS for the (-) position is drawn using a dashed green line. The dichroic signal (difference) is the hatched red area. $\mathbf{G}=(200)$ for bcc-Fe and fcc-Ni and $(100)$ for hcp-Co. \label{fig:dosdich}}
\end{figure*}

Besides $\gamma^{(j,l)}$, the other factors determining the value of $q_z$ are the energy of the edge, i.e.\ the energy lost by the probe electron, the tilt with respect to the zone axis and whether the excited beam is in a HOLZ. These last factors have been included in our calculation. Only the variations due to $\gamma^{(j,l)}$ are neglected, thus giving rise to an error $\lesssim$ 1\%. If a more accurate treatment would be needed, the smooth behavior of MDFF with respect to $q_z$ would allow to use simple linear or quadratic interpolation/extrapolation methods.

As mentioned in the introduction and explained in Refs.~\onlinecite{nature,Hitchcock}, dichroism in the TEM is made possible by the analogous role that the polarization vector $\bm{\varepsilon}$ and the wave vector transfer $\mathbf{q}$ play in the dipole approximation of the DDSCS. However we do not restrict our calculations to the dipole approximation. We use the more complete expression Eq.~(\ref{mdff}).

To evaluate the accuracy of the dipole approximation, we compare the dipole approximation of MDFF with the full calculation (with $\lambda$ up to 3) also showing $\lambda$-diagonal components of the MDFF, Fig.~\ref{fig:dipapprox}. Because the dominant contribution to the signal originates from (dipole allowed) $2p \to 3d$ transitions, the $\lambda=\lambda'=1$ term nearly coincides with the total MDFF. While the dipole approximation works relatively well for the studied systems, particularly the MDFF divided by squares of momentum transfer vectors (right column of the Fig.~\ref{fig:dipapprox}), it has significantly different asymptotic behaviours for larger $\mathbf{q}$-vectors. The $\lambda=\lambda'=1$ term provides a much better approximation, which remains very accurate also in the large $q$ region.
 
It is worth mentioning that thanks to the properties of the Gaunt coefficients the $2p \to 3d$ transitions are all included in the $\lambda=1$ and $\lambda=3$ contributions. Thanks to the negligible value of the radial integrals for $\lambda=3$ the terms with $\lambda=1$ account for the large majority of the calculated signal. The contributions from $\lambda=0,2$ describe transitions from $2p$ to valence $p$ or $f$ states and are always negligible due to the composition of the density of states beyond the Fermi level. They practically overlap with the zero axis in all the six parts of Fig.~\ref{fig:dipapprox}.

It can be shown~\cite{UMlacbed} that in the dipole approximation the real part of the MDFF is proportional to $\mathbf{q} \cdot \mathbf{q'}$ and the imaginary part is proportional to $\mathbf{q} \times \mathbf{q'}$. A little algebra can thus show that the imaginary part of the MDFF is, in the geometry described in the caption of Fig.~\ref{fig:dipapprox}, constant with respect to $q_x$. As expected, the DFF (which is proportional to $q^2$) has a minimum at $q_x = 0$, where $S(\mathbf{q},E)/q^4$ has a maximum. For the MDFF (and corresponding $S(\mathbf{q},\mathbf{q'},E)/q^2 q'^2$) these minima and maxima are centered at $q_x=-G/2=-0.76$ $a.u.^{-1}$ where $|q_x|=|q_x'|$.

\section{Results}

\begin{figure}[t]
  \includegraphics[width=6cm,angle=270]{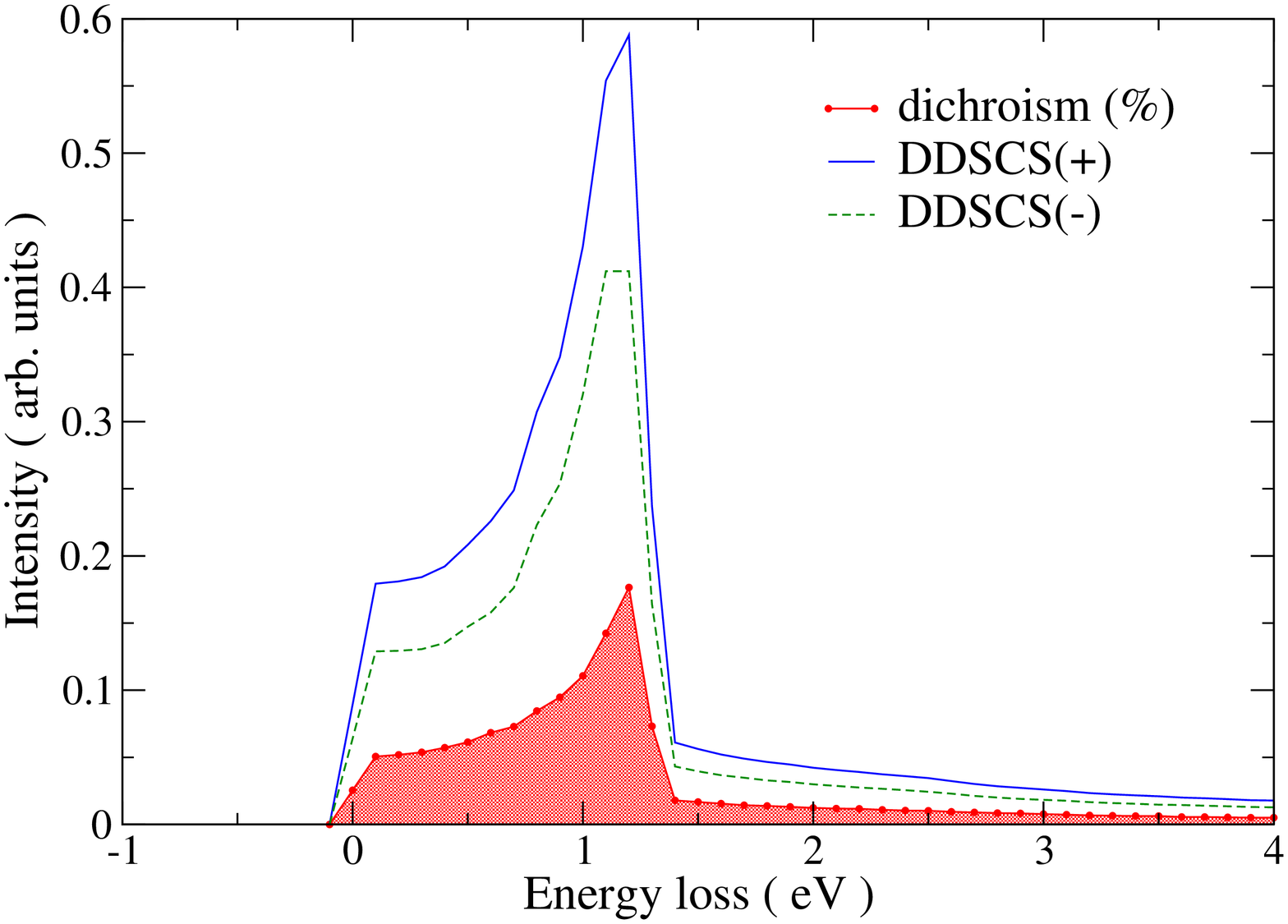}
  \caption{L$_3$ peak of hcp-Co calculated for the $\mathbf{G}=(110)$ systematic row at 18nm.
           See caption of Fig.~\ref{fig:dosdich}. 
           The peaks have been renormalized so that their sum is 1, therefore their difference is the dichroic signal (ca. 15\% in this case).
           \label{fig:Co110}}
\end{figure}

We summarize the results obtained for body-centered cubic iron (bcc-Fe), hexagonal close-packed cobalt (hcp-Co) and face-centered cubic nickel (fcc-Ni) crystals, which are also the first samples prepared for EMCD measurements. These results are valuable for optimization of the experimental setup. 

The geometry setup for observing the dichroic effect~\cite{nature} consists in creating a two-beam case by tilting the beam away from a zone axis (here $(001)$) by a few degrees
and then setting the Laue circle center equal to $\mathbf{G}/2$ for the $\mathbf{G}$ vector to be excited. In analogy to XMCD, where two measurements are performed for left- and righ-handed circularly polarized light, here we perform two measurements by changing the position of the detector, which lies once at the top and once at the bottom of the Thales circle having as diameter the line connecting the diffraction spots $\mathbf{0}$ and $\mathbf{G}$. This geometry setup, together with the crystal structure, is an input for the calculation of the Bloch wave coefficients (within the systematic row approximation) using the dynamical diffraction theory code described in section \ref{sub:ddt}.

The electronic structure was calculated using the WIEN2k package\cite{wien2k}, which is a state-of-the-art implementation of the full-potential linearized augmented plane waves method. The experimental values of lattice parameters were used. More than 10000 $\mathbf{k}$-points were used to achieve a very good converge of the Brillouin zone integrations. Atomic sphere sizes were 2.2, 2.3 and 2.2 bohr radii for bcc-Fe, hcp-Co and fcc-Ni, respectively. The resulting electronic structure was the input for the calculation of the individual MDFFs required for the summation (see Section \ref{sec:mdff}).

In the three studied cases the dichroic effect is dominated by the transitions to the unoccupied $3d$ states. The $d$-resolved spin-up density of states (DOS) is almost fully occupied, while the spin-down $d$-DOS is partially unoccupied. In Fig.~\ref{fig:dosdich} we compare the $d$-DOS with the dichroic signal at the $L_3$ edge. 
Due to negligible orbital moments in these compounds the $L_2$ edge shows a dichroic signal of practically the same magnitude but with opposite sign. The shape of the calculated dichroic peaks corresponds to the difference of spin-up and spin-down $d$-DOS, similarly to XMCD, as it was shown for the same set of systems in Ref.~\onlinecite{xmcdmodel}.
The calculations were performed within systematic row conditions with $\mathbf{G}=(200)$ for bcc-Fe and fcc-Ni and $\mathbf{G}=(100)$ for hcp-Co. The sample thicknesses were set to 20 nm, 10 nm and 8 nm for bcc-Fe, hcp-Co and fcc-Ni, respectively. These values were found to be optimal for these systems in the given experimental geometry.

An interesting point is the comparison of the strength of the dichroic signal. According to the $d$-DOS projections one would expect comparable strength of signals for the three elements under study. But the dichroic signal of hcp-Co seems to be approximately a factor of two smaller than that of the other two. 
The reason for that can be explained by simple geometrical considerations starting from Eq.~(\ref{dscsfin}).
For simplicity we consider only the main contributions: the DFF $S(\mathbf{q},\mathbf{q},E)$
and the MDFF $S(\mathbf{q},\mathbf{q'},E)$ with $\mathbf{q} \perp \mathbf{q'}$.
For bcc-Fe and fcc-Ni the summation over $\mathbf{u}$ within the Bravais cell leads always to the structure 
factor 2 and 4, respectively, because $\mathbf{q'}-\mathbf{q}=\mathbf{G}$ is a kinematically allowed reflection.
This factor cancels out after division by the number of atoms in the Bravais cell.
Therefore it does not matter, what is the value of $\mathbf{q}$-vectors, the sum over the atoms is equal to $S(\mathbf{q},\mathbf{q'},E)/q^2q'^2$ itself. On the other hand, the unit cell of hcp-Co contains two equivalent atoms at positions $\mathbf{u}_1 = (\frac{1}{3},\frac{2}{3},\frac{1}{4})$ and
$\mathbf{u}_2 = (\frac{2}{3},\frac{1}{3},\frac{3}{4})$. For the two DFFs $\mathbf{q}=\mathbf{q'}$
and the exponential reduces to 1; since there are two such terms, after division by $N_\mathbf{u}$ the sum equals
again the DFF itself. But for the main MDFF we have $\mathbf{q} \perp \mathbf{q'}$ and the exponential 
factor will in general weight the terms. One can easily see, that $\mathbf{q'}-\mathbf{q}=\mathbf{G}$.
For the $\mathbf{G}=(100)$ systematic row case, which was used for
calculation of hcp-Co in Fig.~\ref{fig:dosdich} the exponentials evaluate to the complex numbers
$-\frac{1}{2} \pm i\frac{\sqrt{3}}{2}$ and $-\frac{1}{2} \mp i\frac{\sqrt{3}}{2}$ for $\mathbf{u}_1$ and $\mathbf{u}_2$, respectively. Because of symmetry, the MDFFs for both atoms are equal and then the sum $\frac{1}{N_\mathbf{u}}\sum_\mathbf{u}$ leads to a factor $-\frac{1}{2}$ for the MDFF contribution, {\it i.e.}\ the influence of its imaginary part, which is responsible for dichroism, on the DDSCS is reduced by a factor of two.

\begin{figure}[t]
  \includegraphics[width=7cm]{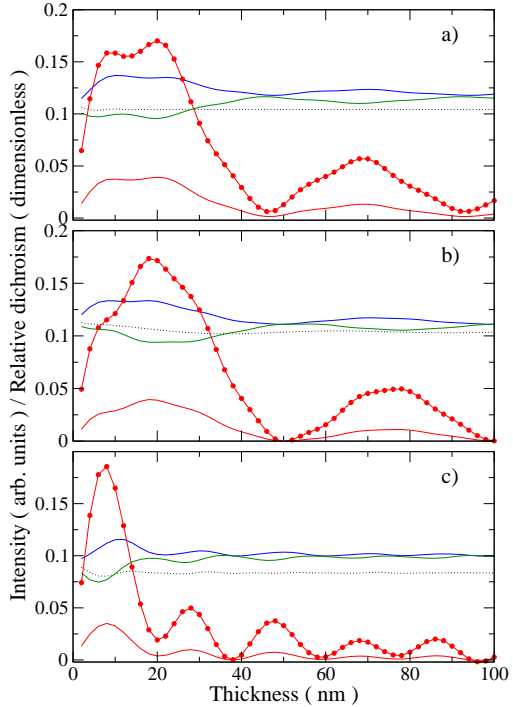}
  \caption{(online color) Dependence of the DDSCS and of the dichroic signal on sample thickness for 
           a) bcc-Fe, b) hcp-Co and c) fcc-Ni. Systematic row vector $\mathbf{G}=(200)$ was used for bcc-Fe and
           fcc-Ni, while for hcp-Co $\mathbf{G}=(110)$ was chosen. 
           The blue and green solid curves are DDSCSs calculated for the (+) and (-) detector positions,
           the dashed black curve is the DFF part of the DDSCS (it is identical for both detector positions).
           The red line with circles is the relative dichroism defined as difference of DDSCSs divided by their sum,
           the red solid curve is the absolute dichroism - difference of DDSCSs.
           \label{fig:thick}}
\end{figure}

\begin{figure}[t]
  \includegraphics[width=7cm,angle=270]{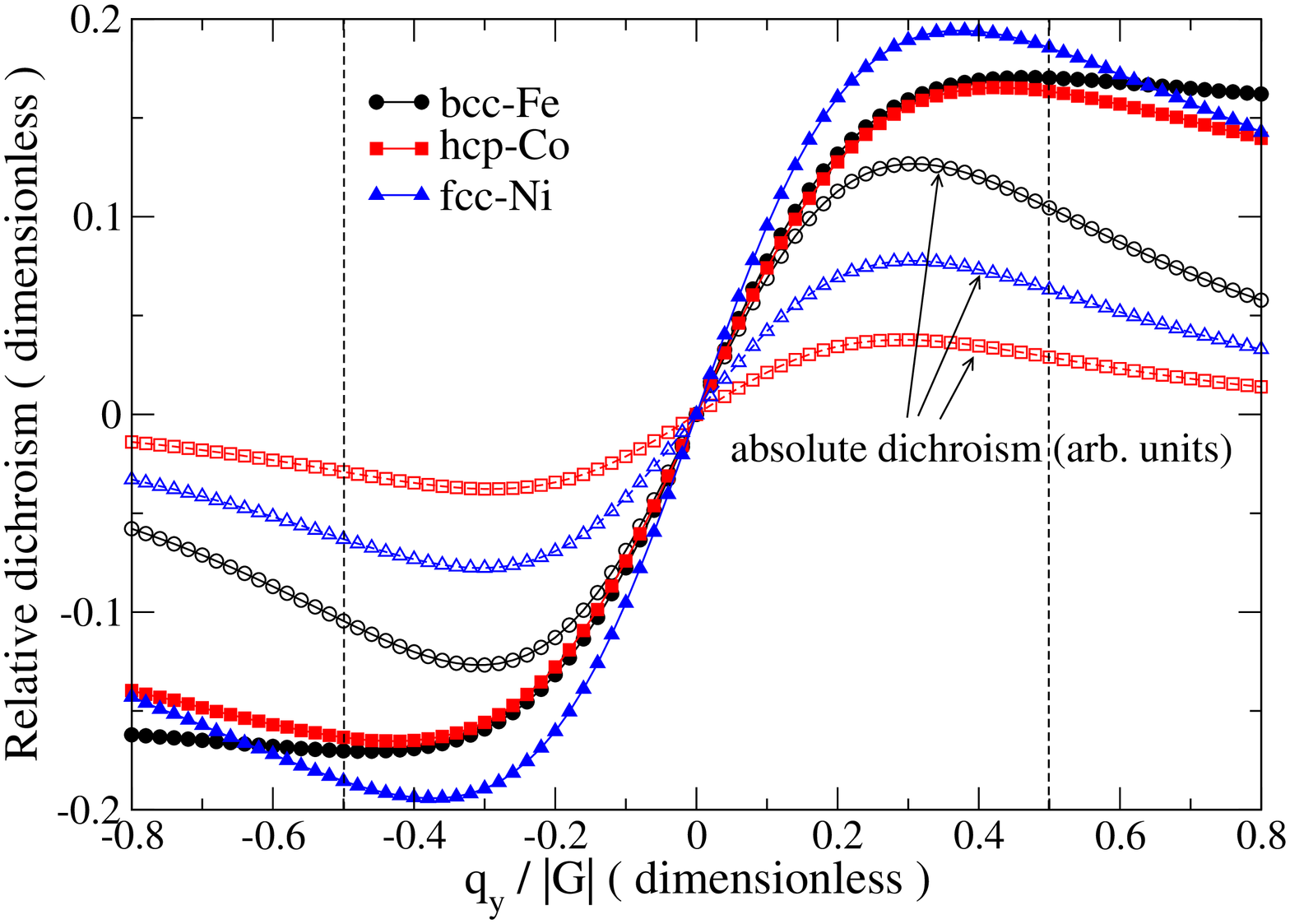}
  \caption{(online color) Dependence of the dichroic signal on detector displacements along $q_y$. 
          The full symbols correspond to the relative dichroic signal while the open symbols to 
           the difference of the DDSCS for both detector positions. These are in arbitrary units and
           their magnitudes are not directly comparable. Vertical lines are showing the \emph{default}
           detector positions.\label{fig:detpos}}
\end{figure}

To optimize the dichroic signal strength of hcp-Co, we require 
$\mathbf{G} \cdot \mathbf{u}_1 = \mathbf{G} \cdot \mathbf{u}_2 = 2 \pi n$, which gives in principle an infinite set of possible $\mathbf{G}$ vectors. The one with lowest $hkl$ indices is $\mathbf{G}=(110)$. A calculation for this geometry setup leads to approximately twice the dichroic signal, see Fig.~\ref{fig:Co110} and compare to the corresponding graph in Fig.~\ref{fig:dosdich}. 

For the optimization of the experimental setup it is important to know how sensitive the results are to variation
of the parameters like the thickness of the sample or the accuracy of the detector position. Another question related to this is also the sensitivity to the finite size of the convergence and collection angles $\alpha$ and $\beta$. In the following text we will address these questions.

The thickness influences the factor $T_{jlj'l'}$ in the Eq.~(\ref{dscsfin}) only. This factor leads to the so called \emph{pendell\"osung} oscillations - modulations of the signal strength as a function of thickness. This also influences the strength of the dichroic signal. Results of such calculations are displayed in Fig.~\ref{fig:thick} (we did not include absorption into these simulations, so that all signal variations are only due to the geometry of the sample). From these simulations it follows that a well defined thickness of the sample is a very important factor. Relatively small variations of the thickness can induce large changes in the dichroic signal, particularly in fcc-Ni. From the figure one can deduce that the optimal thickness for a bcc-Fe sample should be between 8 nm and 22 nm (of course, due to absorption, 
thinner samples within this range would have a stronger signal), for hcp-Co between 15 nm and 22 nm and for fcc-Ni it is a relatively narrow interval - between 6 nm and 10 nm. However, we stress that these results depend on the choice of the systematic row vector $\mathbf{G}$. For example hcp-Co with $\mathbf{G}=(100)$ (instead of $(110)$ shown in Fig.~\ref{fig:thick}) has a maximum between 5 nm and 15 nm (although it is much lower, as discussed before).

Taking the optimal thickness, namely 20 nm, 18 nm and 8 nm for bcc-Fe, hcp-Co and fcc-Ni, respectively, we calculated the dependence of the dichroic signal on the detector position. We particularly tested changes of the dichroic signal when the detector is moved away from its \emph{default} position in the direction perpendicular to $\mathbf{G}$, see Fig.~\ref{fig:detpos}. It is interesting to note that the maximum absolute difference occurs for a value of $q_y$ smaller than $|G|/2$. This can be qualitatively explained by considering the non-zero value of $q_z$ and $q'_z$, {\it i.e.} $\mathbf{q}$ and $\mathbf{q'}$
are not exactly perpendicular at the default detector positions. Moreover the MDFF enters the summation always divided by $q^2q'^2$ and the lengths of $\mathbf{q}$-vectors decrease with decreasing $q_y$.
The important message we can deduce from this figure is that the dichroic signal is only weakly sensitive to the accuracy of $q_y$ since even displacement by 10-20\% in the detector default $q_y$ positions ({\it i.e.} $q_y = \pm G/2$) do not affect significantly the measured dichroic signal.

Related to this is a study of the dependence of the dichroic signal on the finite size of the convergence and collection angles $\alpha$ and $\beta$. We performed a calculation for the three studied metals and found that collection and convergence half-angles up to 2 mrad weakens the relative dichroic signal by less than 10\%.

\section{Conclusions}

We have developed a computer code package for the calculation of electron energy loss near edge spectra, which includes the theory of dynamical Bragg diffraction. We applied the code to the recently discovered phenomenon of magnetic chiral dichroism in the TEM and we demonstrated the relation of the dichroic peak shape to the difference of $d$-projections of the spin-resolved density of states in analogy with similar observation for XMCD.

Using this code we examined the validity of the dipole approximation, which is often assumed. We found that for the  $3d$ ferromagnetic systems studied it is a reasonable approximation, however with wrong asymptotic properties - it overestimates the contributions from larger $\mathbf{q}$-vectors. A very accurate approximation for the studied systems is the $\lambda=\lambda'=1$ approximation, which treats  appropriately the dominant $p \to d$ dipole transitions and remains very accurate also for large $q,q'$.

In order to provide guidance to the experimentalist we have investigated the strength of the dichroic signal as a function of the sample thickness and the precision of the detector placement. While the dichroic signal strength is rather robust with respect to the precision of the detector placement, 
the thickness of the specimen influences the signal considerably. Therefore it might be a challenge to produce samples with optimum thickness and selecting the best systematic row Bragg spot. Our calculations yield best thicknesses in order to detect EMCD of the iron and nickel samples for the systematic row $\mathbf{G}=(200)$ 
to be 8-22 nm and 6-10 nm, respectively, and for cobalt in the systematic row $\mathbf{G}=(110)$ to be 15-22 nm.

\begin{acknowledgments}
We thank Dr.\ C\'{e}cile H\'{e}bert and Dr.\ Pavel Nov\'{a}k for stimulating discussions. This work has been supported by the European Commission, contract nr. 508971 (CHIRALTEM).
\end{acknowledgments}


\begin{thebibliography}{99}
\bibitem{vgnote} The WIEN2k package can calculate X-ray structure factors. By supplying the potential instead of charge density it is possible to use the same code to calculate the electron structure factors using a simple script.
\end{thebibliography}
\end{document}